# Is human atrial fibrillation stochastic or deterministic?—Insights from missing ordinal patterns and causal entropy-complexity plane analysis

Konstantinos N. Aronis, Ronald D. Berger, Hugh Calkins, Jonathan Chrispin, Joseph E. Marine, David D. Spragg, Susumu Tao, Harikrishna Tandri, and Hiroshi Ashikaga

# Is human atrial fibrillation stochastic or deterministic?—Insights from missing ordinal patterns and causal entropy-complexity plane analysis

Konstantinos N. Aronis,[1] Ronald D. Berger,[1] Hugh Calkins,[1] Jonathan Chrispin,[1] Joseph E. Marine,[1] David D. Spragg,[1] Susumu Tao,[1] Harikrishna Tandri,[1] and Hiroshi Ashikaga[1,2,a)]

[1]*Cardiac Arrhythmia Service, Johns Hopkins University School of Medicine, Baltimore, Maryland 21287, USA*
[2]*IHU Liryc, Electrophysiology and Heart Modeling Institute, Fondation Bordeaux Université, F-33600 Pessac-Bordeaux, France*

The mechanism of atrial fibrillation (AF) maintenance in humans is yet to be determined. It remains controversial whether cardiac fibrillatory dynamics are the result of a deterministic or a stochastic process. Traditional methods to differentiate deterministic from stochastic processes have several limitations and are not reliably applied to short and noisy data obtained during clinical studies. The appearance of missing ordinal patterns (MOPs) using the Bandt-Pompe (BP) symbolization is indicative of deterministic dynamics and is robust to brief time series and experimental noise. Our aim was to evaluate whether human AF dynamics is the result of a stochastic or a deterministic process. We used 38 intracardiac atrial electrograms during AF from the coronary sinus of 10 patients undergoing catheter ablation of AF. We extracted the intervals between consecutive atrial depolarizations (AA interval) and converted the AA interval time series to their BP symbolic representation (embedding dimension 5, time delay 1). We generated 40 iterative amplitude-adjusted, Fourier-transform (IAAFT) surrogate data for each of the AA time series. IAAFT surrogates have the same frequency spectrum, autocorrelation, and probability distribution with the original time series. Using the BP symbolization, we compared the number of MOPs and the rate of MOP decay in the first 1000 timepoints of the original time series with that of the surrogate data. We calculated permutation entropy and permutation statistical complexity and represented each time series on the causal entropy-complexity plane. We demonstrated that (a) the number of MOPs in human AF is significantly higher compared to the surrogate data ($2.7 \pm 1.18$ vs. $0.39 \pm 0.28$, $p < 0.001$); (b) the median rate of MOP decay in human AF was significantly lower compared with the surrogate data ($6.58 \times 10^{-3}$ vs. $7.79 \times 10^{-3}$, $p < 0.001$); and (c) 81.6% of the individual recordings had a rate of decay lower than the 95% confidence intervals of their corresponding surrogates. On the causal entropy-complexity plane, human AF lay on the deterministic part of the plane that was located above the trajectory of fractional Brownian motion with different Hurst exponents on the plane. This analysis demonstrates that human AF dynamics does not arise from a rescaled linear stochastic process or a fractional noise, but either a deterministic or a nonlinear stochastic process. Our results justify the development and application of mathematical analysis and modeling tools to enable predictive control of human AF. *Published by AIP Publishing.*
https://doi.org/10.1063/1.5023588

**Atrial fibrillation (AF) is the most common cardiac arrhythmia in humans and is associated with significant morbidity and mortality. The current standard of care includes interventional catheter ablation in selected patients, but the success rate is limited. The major limitation of the current approach to AF is the lack of fundamental understanding of its underlying mechanism. Specifically, it remains unclear whether human AF dynamics is a deterministic or a stochastic process. Here, we assessed for determinism in human AF by evaluating the properties of the symbolic representation of intracardiac electrical recordings obtained from patients. Specifically, we evaluated (a) the number of missing ordinal patterns (MOP); (b) the rate of missing ordinal pattern decay for the increased length of the time series; and (c) the causal-entropy complexity plane of the Bandt-Pompe (BP) symbolic representation. When used together, these are powerful tools to detect determinism, even in the presence of experimental noise and brief time series. We demonstrate that AF dynamics cannot be modeled as a rescaled linear stochastic process or fractional noise. Consequently, AF dynamics arise from either a deterministic or a nonlinear stochastic process.**

## I. INTRODUCTION

Atrial fibrillation (AF) is the most common cardiac arrhythmia in humans, with an increasing prevalence that is estimated to rise to 12.1 million in 2030 in the United States and a significant morbidity and mortality associated with it.[1] AF is characterized by an "irregularly irregular" heart rhythm

a)Author to whom correspondence should be addressed: hashika1@jhmi.edu. Telephone: 410-955-7534. Fax: 443-873-5019.

and a seemingly disorganized activation of the left and right atrium.[2] The current therapeutic approach to AF using interventional catheter ablation has modest efficacy, with recurrence rates up to ∼30%.[3–5] The main reason for these limited clinical outcomes is that, despite the advancement in mapping and catheter ablation technology, the mechanisms of AF maintenance in human are not well-understood. For example, to describe human AF dynamics, both deterministic[6–10] and stochastic[11,12] models have been developed. It remains controversial whether the human AF dynamics result from a deterministic or a stochastic process.[13–16]

Elucidating whether the disorganized dynamics observed in human AF is the result of a deterministic or a stochastic process is essential for a proper physical description of AF. Identifying determinism in AF time series is critically important for understanding the mechanism of modeling and predicting AF. Dynamics arising from deterministic processes can be described with relatively few non-linear modes, while dynamics arising from stochastic processes are better described by statistical approaches. Deterministic dynamics is predictable on relatively short time scales and might form stable attracting patterns in the phase space, while stochastic processes are random at any time step and do not form attractors. Discrimination between deterministic and stochastic dynamics can be extremely challenging, especially when the time series under investigation are contaminated with experimental noise, since both processes share many features.[17] Traditional methods for detecting deterministic chaos such as the correlation dimension,[18] Kolmogorov entropy,[19] Lyapunov exponents,[20] nonlinear forecasting models,[21] determinism test,[22] noise titration,[23] and 0–1 test[24] are not readily applicable to biomedical recordings as they are sensitive to experimental noise, require long and/or stationary time series, and/or are sensitive to initial parameter selection. Furthermore, these tests are not fully reliable and have several limitations.[25–32]

Symbolic representation of theoretical and experimentally acquired time series, with ordinal patterns using the Bandt and Pompe's (BP) methodology, has given a new insight into time series characterization and detection of determinism.[33] The emergence of ordinal patterns that never appear in a time series of adequate length (*"forbidden ordinal patterns"* or FOP) distinguishes deterministic processes from uncorrelated stochastic processes.[33–36] Amigó et al. demonstrated that the decay rate of the missing ordinal patterns (MOP) as a function of the time series length can be used to distinguish deterministic from stochastic processes in relatively short and noisy time series.[34,35,37] The term MOP over FOP is preferred in analysis of time series contaminated with experimental noise, since all ordinal patterns will eventually emerge if these time series are of adequate length and thus are not truly "forbidden" but "missing" in a specific time series length segment. Furthermore, the analysis of the BP symbolic representation has been extended by linking it to the causal entropy-complexity plane. Calculation of permutation entropy and permutation statistical complexity of the BP symbolic representation of the time series under investigation, and representation of the results on the causal entropy-complexity plane, is a powerful tool for detection of determinism.[17] The causal entropy-complexity plane can discriminate deterministic series contaminated with correlated noise from pure noise with long-term correlations.[38] A limitation of the MOP paradigm and the causal entropy-complexity plane analysis is that they have not been validated in discriminating deterministic from nonlinear stochastic processes.

The aim of this study was to assess whether human AF dynamics is the result of a deterministic or a stochastic process using the Bandt-Pompe symbolization and assessing MOP and the symbolic time series representation on the entropy-complexity plane. Our hypothesis was that (1) the BP symbolization and causal entropy-complexity plane analysis can be applied to clinically acquired electrograms of human AF to provide insights into the underlying dynamics; and (2) Human AF is not the result of a rescaled linear stochastic process or a fractional noise. To test our hypotheses, we recorded intracardiac bipolar atrial electrograms of AF from the coronary sinus of patients referred for catheter ablation of AF. We extracted the intervals between consecutive atrial depolarizations (AA interval) as representative of the local atrial macroscopic dynamics. For each AA time series, we generated 40 iterative amplitude-adjusted, Fourier-transform (IAAFT) surrogate time series that have the same frequency spectrum, autocorrelation, and probability distribution with the original time series. We constructed the BP symbolic representation of the AA interval time series and the surrogate data and compared the number of MOPs and the rate of MOP decay. The null hypothesis is that the AA time series is a rescaled Gaussian linear stochastic process, and thus, the number of MOPs and rate of MOP decay will be the same between AA time series and surrogate data. If the null hypothesis is rejected, then the system is deterministic or nonlinear stochastic. We also calculated permutation entropy and permutation statistical complexity of the BP symbolic representation of AA time series and surrogate data and plotted the results on the causal entropy-complexity plane.

## II. METHODS

### A. Intracardiac recordings

We enrolled 10 patients who were referred for a standard catheter ablation for symptomatic, drug-refractory AF at the Johns Hopkins Hospital between August 2017 and October 2017. The protocol was approved by the Johns Hopkins Medicine Institutional Review Board and all participants provided written informed consent. All patients underwent pre-procedural transesophageal echocardiogram to rule out intracardiac thrombus. We introduced a 5-Fr decapolar catheter (Dynamic Tip 2-5-2 Boston Scientific, Marlborough, MA; inter-electrode distance 2 mm between poles and 5 mm between bipolar pairs) in the right femoral vein and advanced it to the coronary sinus. In one patient who presented in sinus rhythm, we induced AF by atrial burst pacing. Induced AF was recorded after >15 min.[39] We recorded intracardiac bipolar electrograms from the decapolar catheter during AF using 3–5 pairs of two immediately adjacent electrodes at the sampling frequency of 977 Hz for the duration of 1.88–3.26 min by the standard clinical electrophysiology recording system (CardioLab, GE Healthcare, Waukesha, WI). We processed each bipolar signal separately. We also simultaneously recorded the surface 12-lead electrocardiogram. We filtered the recorded time series and removed the ventricular signals as previously described (Fig. 1).[40] We excluded from analysis the recordings with high levels of noise that preclude visual identification of atrial signals. Finally, we included a total of 38 intracardiac recordings, as adjudicated by two clinical cardiac electrophysiologists. We defined the atrial depolarization as any peak exceeding 0.02 mV in amplitude located at least 102 ms away from the prior peak, which was shorter than the atrial effective refractory period (AERP) in all cases. We confirmed the accurate identification of atrial depolarization with visual inspection of all detected peaks. If necessary, we made adjustments on the peak detection thresholds to ensure accurate identification of atrial depolarization. We defined the AA interval as the time interval between two consecutive atrial depolarizations [Fig. 2(a)].

### B. Surrogate data

The BP symbolization has been used in an IAAFT surrogate data framework for the detection of non-linear determinism in both theoretical and experimental time series. The presence of a higher number of MOPs in the time series under examination compared to IAAFT surrogate data distinguishes deterministic time series from correlated stochastic time series.[41] IAAFT surrogate data are generated such that the surrogate data have the same power spectrum, autocorrelations, and probability distribution with the original time series. As a result, the derived surrogates have the same probability distribution and power spectrum with potential high order correlations being randomized.[42,43] In the present study, we used IAAFT surrogate data created with the improved algorithm of Schreiber and Schmitz,[43] using the implementation by Leontitsis *et al.* (www.mathworks.com/matlabcentral/fileexchange/1597). For each AA time series, we generated 40 IAAFT surrogate time series. When working with the IAAFT surrogate data, the null hypothesis is that the original time series under investigation is a rescaled Gaussian linear stochastic process. If the null hypothesis is rejected, then the system is deterministic or nonlinear stochastic with probability of falsely rejecting the null hypothesis equal to the p-value of the statistical test used for hypothesis testing.

### C. The Bandt-Pompe symbolization and the Amigó process for detection of determinism

To generate ordinal patterns from the intervals between atrial depolarizations (AA time series), we used the BP methodology.[33–37] The details of the BP methodology are described in supplementary materials (Appendix A). Briefly, a time series $X$ was mapped to a vector of length equal to the embedding dimension $D$ ($D \in \mathbb{N}$), which contains elements of the time series delayed by $\tau$ ($\tau \in \mathbb{N}$). The elements of each vector were replaced by their rank in the vector. This new vector is an observed ordinal pattern for the time series [Fig. 2(a)]. There are $D!$ possible ordinal patterns for a time series, and for $D = 5$, all possible ordinal patterns are depicted in Fig. 2. Ordinal patterns that have not appeared in a length $L$ of time series are called missing ordinal patterns (MOP).[34,35,44]

Amigó *et al.* demonstrated that the rate of decay of MOP for increasing $L$ can be used to discriminate deterministic time series contaminated with noise from pure uncorrelated noise. Specifically, the number of MOPs decays exponentially with increasing $L$ and the rate of decay is significantly different from that of uncorrelated stochastic processes.[35,37] The MOP paradigm has been used to successfully detect determinism in high dimensional dynamics.[36] MOPs can be used to detect determinism even in the setting of irregular-sampling, missing data, and timing jitter.[45] MOPs have been used to detect determinism in financial time series[46,47] and epileptic brain states.[48–50] A limitation of this method is the inability to differentiate deterministic from stochastic processes that exhibit long-term correlations and nonlinear stochastic processes. Specifically, the persistence of MOP is not necessarily a signature of underlying determinism, because the same persistence is found in stochastic time series with long term correlation structures.[44] To overcome this limitation, Kulp *et al.* used IAAFT surrogate data and the BP MOP paradigm, to detect non-linear determinism in both theoretical and experimental deterministic time series and distinguish them from correlated stochastic time series.[41] There are no studies to date evaluating the ability of the MOP paradigm (as a stand-alone method or in a surrogate data framework) in distinguishing determinism from nonlinear stochastic processes.

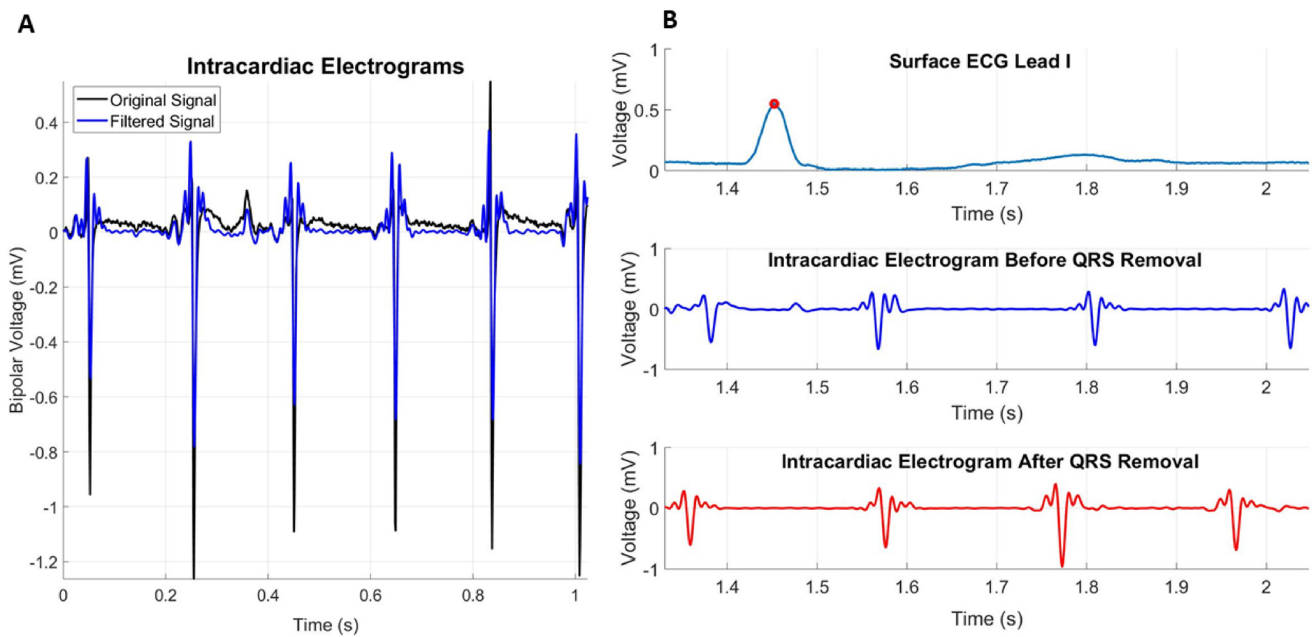

FIG. 1. Example of intracardiac electrograms signal recorded. Panel (a): snapshot of the recorded electrical signal. The black signal shows the raw signal as recorded with CardioLab. The blue signal shows the signal after high-pass and low-pass filtering. Panel (b): example of removal of ventricular depolarization from intracardiac electrogram. The first (upper) picture shows the surface electrocardiographic signal recorded at the body surface. The red circle shows the peak of the QRS complex. The QRS complex represents ventricular depolarization. The second (middle) picture represents the intracardiac atrial electrogram before QRS subtraction and the third (lower) picture represents the intracardiac electrograms after QRS subtraction. Notice the small bump the first atrial depolarization that gets removed after QRS subtraction.

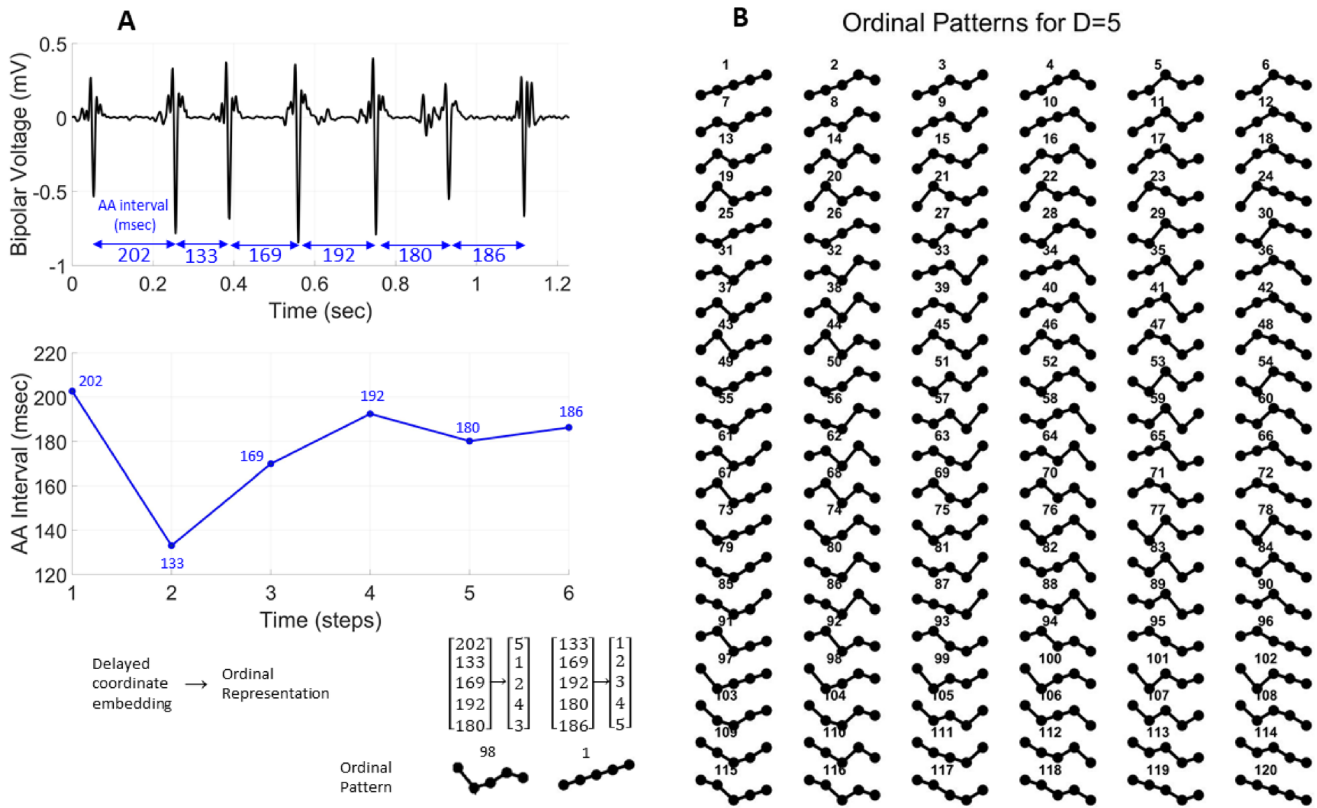

FIG. 2. Example of derivation of AA time series and their symbolic representation. Panel (a): The upper signal represents the first 1.2 s of an intracardiac recording after filtering and ventricular depolarization removal as described in Fig. 1. Each spike is an atrial depolarization. The AA interval is the time interval between consecutive atrial depolarization. The 6 AA intervals between the first 7 atrial depolarizations of this recording are noted in ms (blue labels). The AA time series resulting from this signal are graphically depicted on the lower diagram. The column vectors below the 5th and 6th AA interval are created with delayed coordinate embedding (D = 5 and $\tau = 1$) and their ordinal representation is provided to the right of these vectors. The delayed-coordinate embedded AA interval of step 5 corresponds to ordinal pattern #98 and the delayed-coordinate embedded AA interval of step 6 corresponds to ordinal pattern #1 according to the BP symbolization. Panel (b): All potential ordinal patterns for $D = 5$ according to the BP symbolization.

In the present study, we used the first 1000 time steps (= 1000 atrial depolarizations) of the AA time series and surrogate data, to ensure that the analyzed time series will be of the same length. We used $D = 5$ and $\tau = 1$ to ensure adequate sampling of the time series $[1000 > (5 + 1)! = 720]$. We calculate the number of MOPs. Subsequently, for each time series, we calculated the rate of MOP decay for increasing length L ($5 \leq L \leq 1000$) by fitting an exponential function of the type

$$MOP(L) = MOP_0 e^{-bL}, \quad 5 \leq L \leq 1000. \quad (1)$$

Here, $MOP(L)$ is the number of MOPs at time series length $L$, $MOP_0$ is the number of missing patterns at L = 5, and $b$ is the time constant for the exponential decay and represents the decay of MOPs. Smaller values of $b$ mean slower decay and are consistent with deterministic over stochastic time series.[35,37] We compared the median number of MOPs and the median rate of MOP decay in the AA time series and surrogate data using the Mann-Whitney U statistical test. We compared the mean number of MOPs and mean rate of MOP decay in the AA time series and surrogate data using a Student's t-test with Welch correction for unequal variances. If the number of MOPs is higher and the rate of MOP decay is lower in the AA time series compared to surrogate data, we then reject the null hypothesis that the AA time series is a rescaled Gaussian linear stochastic process, and therefore, the AA time series is deterministic or nonlinear stochastic.[41] The probability of falsely rejecting the null hypothesis is equal to the two-sided p-value derived from the Man-Whitney U test and Student's t-test. Furthermore, we compared the rate of MOP decay of each individual time series with the mean ±95% confidence interval of the rate of MOP decay of the 40 corresponding surrogate data. If the rate of decay of the individual time series is lower than the 95% confidence interval band, then the specific time series is deterministic or nonlinear stochastic.

### D. Causal entropy-complexity plane

Each time series $X$ is represented on the causal entropy-complexity plane $\mathcal{H}[P] \times \mathcal{C}_{JS}[P]$ as a point $(\mathcal{H}[P], \mathcal{C}_{JS}[P])$.[17,38] Here, $\mathcal{H}[P]$ is the permutation entropy and $\mathcal{C}_{JS}[P]$ the permutation statistical complexity of the probability distribution $P$ of the observed ordinal patterns $\pi$ in the time series $X$.[17,38] Calculation of permutation entropy and permutation statistical complexity are described in detail in *supplementary materials (Appendix B)*. Shannon entropy-based measures quantify the information content, or uncertainty, associated with the physical process described by $P$[51] but do not quantify the degree of structure or patterns of the process.[52] Measures of statistical complexity are necessary to capture the organizational properties of the process,[53] to detect essential details of the dynamics, and to differentiate different degrees of periodicity and chaos.[54] The BP symbolization takes into account the time-causality in the derivation of the probability distribution $P$ associated with the time series under investigation.[33] Because the probability distribution is derived from the BP methodology, all the advantages associated with it, including simplicity, low computational cost, robustness, and invariance with respect to monotonous transformations, are inherited by the $\mathcal{H}[P] \times \mathcal{C}_{JS}[P]$ plane analysis.

The causal entropy-complexity plane of the BP symbolic representation of a time series has been used to detect determinism in time series[17] and to distinguish deterministic time series contaminated with correlated noise from purely correlated noise.[38] Specifically, deterministic time series, even if contaminated with correlated noise (of various intensities and strengths of correlation), maintain higher complexity levels for the same entropy levels on the $\mathcal{H}[P] \times \mathcal{C}_{JS}[P]$ plane.[38] Fractional Brownian motion (fBm) is a Gaussian process that starts at zero, has an expected value of zero at any timepoint, and has a covariance structure between two timepoints $t$ and $s$ described by the following equation:

$$\mathbb{E}\left(B_t^H, B_s^H\right) = \frac{1}{2}\left(s^{2H} + t^{2H} - |t-s|^{2H}\right). \quad (2)$$

Here, $B_t^H$ denotes fBm with Hurst parameter $H \in (0,1)$ at timepoint $t$ and $\mathbb{E}\left(B_t^H, B_s^H\right)$ the covariance of the process between timepoints $t$ and $s$.[55] If $H = 1/2$, then the increments of the process are not correlated, and the process is in fact a Brownian motion or Wiener process; if $H > 1/2$, then the increments of the process are positively correlated; if $H < 1/2$, then the increments of the process are negatively correlated. The trajectory of fractional Brownian motion with Hurst exponents $0 < H < 1$ on the $\mathcal{H}[P] \times \mathcal{C}_{JS}[P]$ plane has been used to split the plane into two areas. Time series that have a trajectory sufficiently above the trajectory of fBm are characterized as deterministic and those with a trajectory below the trajectory of fBm are characterized as stochastic.[56,57] Similar to the BP paradigm, a limitation of this approach is that it has not been validated in distinguishing determinism from nonlinear stochastic processes.

In the present study, we estimated the permutation entropy and permutation statistical complexity for each AA time series, surrogate data, and 200 synthetic fBm time series with Hurst parameters increasing from 0.1 to 1 with a step of 0.1 (20 time series per for each exponent). We plotted each time series as a point on the $\mathcal{H}[P] \times \mathcal{C}_{JS}[P]$ plane. IAAFT surrogate data have by design the same probability distribution with the original time series and are thus expected to lie on the same trajectory on the $\mathcal{H}[P] \times \mathcal{C}_{JS}[P]$ plane with the AA time series. To statistically compare the position of the AA time series relative to that of fBm on the $\mathcal{H}[P] \times \mathcal{C}_{JS}[P]$ plane, we fit a separate cubic regression curve to the points derived from the AA time series and the points derived from fBm using a least squares technique. We constrained the fitted cubic curve to pass from the point (0,0). We calculated the 99% confidence intervals for each curve. If the fitted curve of the AA time series lies at a higher complexity trajectory compared to the fitted curve of fBm, with non-overlapping confidence intervals, then AA time series is deterministic or nonlinear stochastic.[56,57] The probability of falsely rejecting the null hypothesis is equal to 1% since we use the 99% confidence intervals for statistical inference. The goodness-of-fit of the linear regression was assessed with the coefficient of determination of the regression model. The coefficient of determination is the proportion of the variance of $\mathcal{C}_{JS}[P]$ that is predictable from $\mathcal{H}[P]$. A coefficient of determination of >90% indicates that the model fits the data very well and a coefficient of determination of 100% indicates that the model fits the data perfectly. Furthermore, we evaluated the accurate discrimination between fBm and AA time series using $\mathcal{H}[P] \times \mathcal{C}_{JS}[P]$ coordinates using a support vector machine approach (*supplementary materials, Appendix C*).

## III. RESULTS

### A. Intracardiac recordings and surrogate data

Figure 3 shows the intracardiac bipolar electrograms, the AA time series, and the corresponding surrogate data obtained from three patients. Each atrial depolarization in the bipolar electrograms contained 2–4 sharp, high-frequency deflections, although the exact morphology of the atrial electrograms varied among patients, leads, and atrial depolarizations. The bipolar electrograms contained the electrical activity of various frequencies and amplitudes, which represents local fragmented electrical activity, far-field signals, and/or noise. The dynamics of atrial depolarization are inherently passed on to the dynamics of the AA time series.[58] Overall, the AA time series showed the stereotypical irregular interval behavior of AF (described clinically as "*irregularly irregular*"). The length of the AA time series was 1026–1297 time steps. We used only the first 1000 time steps for analysis to allow comparative assessments among different recordings. The mean (±standard deviation) AA interval of the original AA time series was 168.4 ± 32 ms (25th–75th percentile was 150.5–183.2 ms). The mean AA interval of the IAAFT surrogate data was 168.2 ± 31 ms (25th–75th percentile was 151.5–182.2 ms), which was similar to that of the original AA time series. This is consistent with the IAAFT surrogate data design described above.

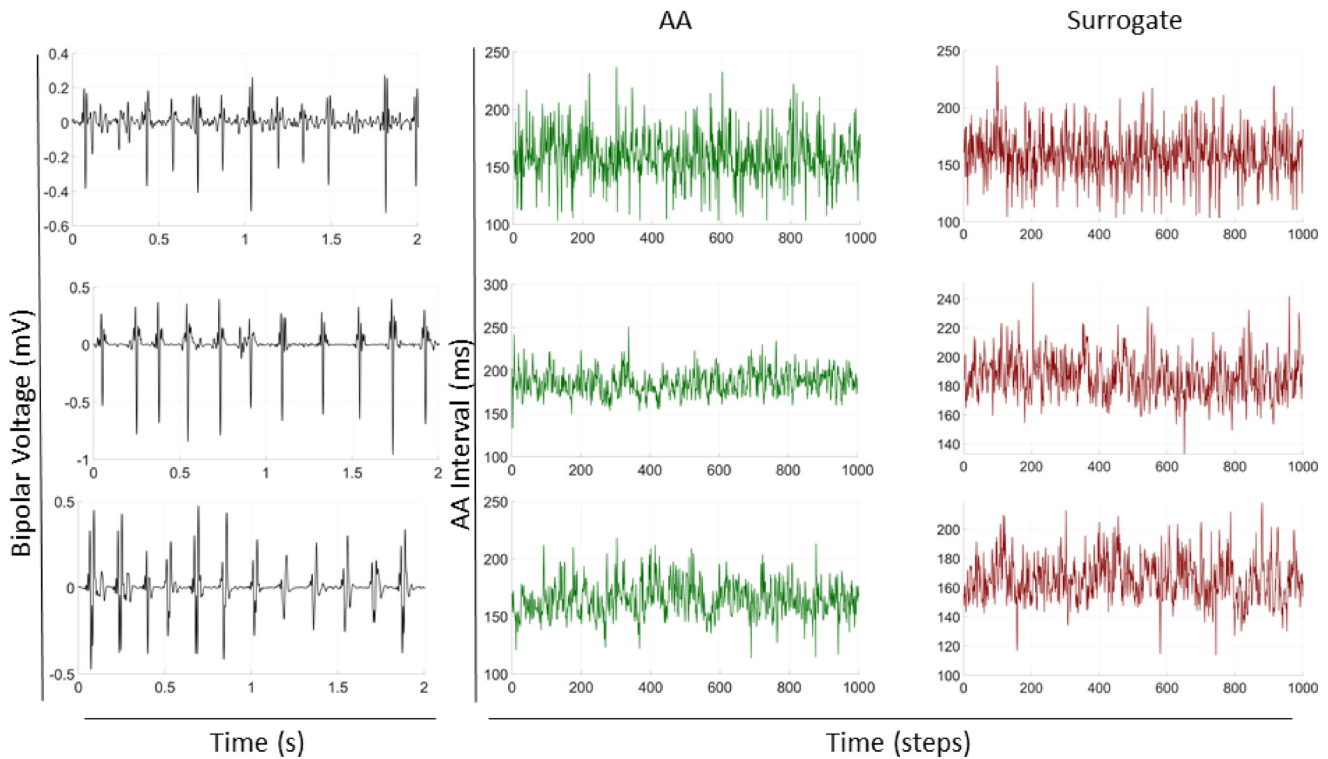

FIG. 3. Examples of processed intracardiac recordings, corresponding AA time series, and surrogate data from 3 patients. Each row represents the bipolar recording from a different patient. The first column shows the first 2 s of the intracardiac bipolar electrograms. The second column shows the AA interval of the first 1000 atrial depolarizations (dark green). The third column shows the IAAFT surrogate data derived from the AA series (dark red).

### B. Missing ordinal patterns and rate of missing ordinal pattern decay

The mean ($\pm$standard error of the mean), the median, and the range of the number of MOPs in the AA time series were $2.8 \pm 1.18$, 1, and 0–41 (25th-75th percentile range: 0–2), respectively. The number of MOPs in the AA time series was significantly higher than the number of MOPs in the IAAFT surrogate data that had a mean of $0.39 \pm 0.28$, a median of 0, and a range of 0 to 12 (25th–75th percentile range: 0 to 0). A box plot comparing the number of MOPs for the AA time series and the IAAFT surrogate data is shown in Fig. 4. The p-value from the Mann-Whitney U comparing the number of MOPs in the AA time series and the IAAFT surrogate data was $<0.001$ The p-value from the Student's t-test comparing the mean number of MOPs between the AA time series and IAAFT surrogate data was $<0.001$. This indicates that the null hypothesis was rejected, and therefore, the AA time series is a result of a deterministic or a nonlinear stochastic process with 99% confidence.

The mean ($\pm$standard error of the mean), the median, and the 25th–75th percentile range of the time constant of MOP decay of the AA time series as defined by (1) were $6.39 \times 10^{-3} \pm 0.31 \times 10^{-3}$, $6.58 \times 10^{-3}$, and $5.33 \times 10^{-3}$ to $7.88 \times 10^{-3}$, respectively. The time constant of MOP decay in the AA time series was significantly lower than the time constant of MOP decay of the IAAFT surrogate data that had a mean of $7.77 \times 10^{-3} \pm 0.03 \times 10^{-3}$, a median of $7.95 \times 10^{-3}$, and a

25th–75th percentile range of $7.16 \times 10^{-3}$ to $8.61 \times 10^{-3}$. An example of the MOP decay with increasing time series length is shown in Fig. 5(a). Box plots comparing the time constant of MOP decay for the AA time series and the IAAFT surrogate data are shown in Fig. 5(b). The p-value from the Mann-Whitney U comparing the time constant of MOP decay in the AA time series and the IAAFT surrogate data was <0.001. The p-value from the Student's t-test comparing the mean rate of decay between AA time series and IAAFT surrogate data was < 0.001. This indicates that the null hypothesis was rejected, and therefore, the AA time series is a result of a deterministic or nonlinear stochastic process with >99% confidence.

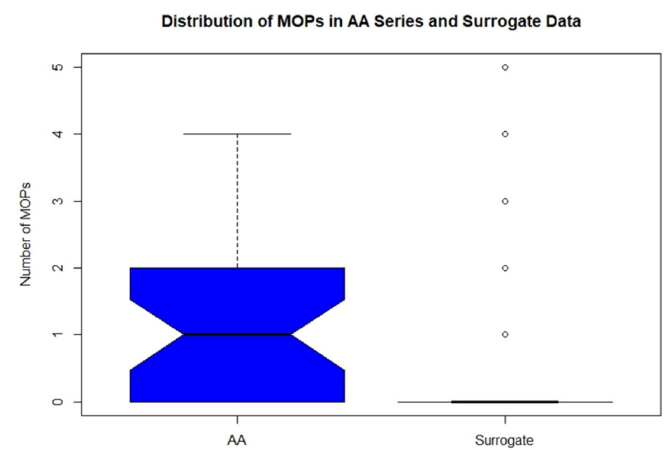

FIG. 4. Box-plot diagrams comparing the number of missing ordinal patterns that emerge in the experimental AA time series data (blue) and the surrogate data (red). The black line in the center of the box represents the median. The notch represents 95% confidence interval of the median. The upper edge of the box represents the 75th percentile and the lower edge of the box represents the 25th percentile. The upper whisker adds 1.5 times the inter-quartile range to the 75th percentile and the lower whisker subtracts 1.5 times the interquartile range from the 25th percentile. Circles denote possible outliers. The median of the number of missing ordinal patterns in the experimental time series is lower than that of the median value of the surrogate data (Mann-Whitney test, $p < 0.001$), suggesting that the null hypothesis of AF dynamics being the result of a rescaled linear stochastic process is rejected.

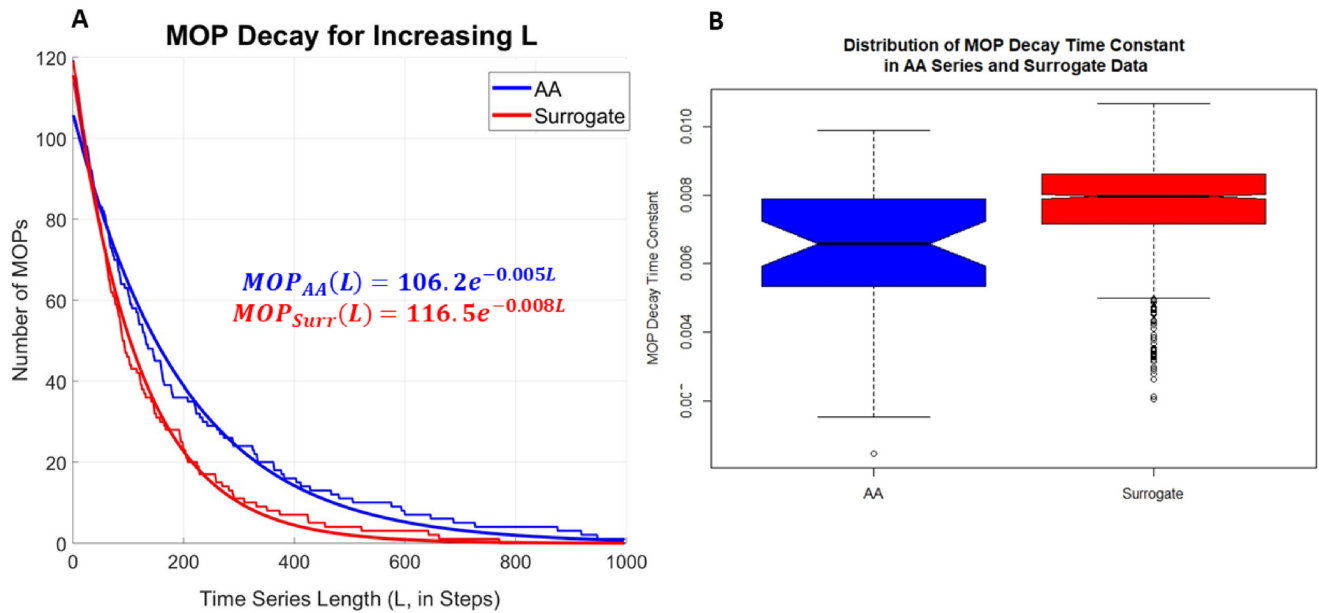

FIG. 5. Panel (a): Example of the exponential decay of the number of missing ordinal patterns with the increasing length of the time series. This graph is created from data from one recording. Experimental AA time series are shown in blue, and surrogate data are shown with red. The solid smooth lines represent the fitted lines, while the stepped lines represent the actual number of missing ordinal patterns for increasing L. The equations represent the equations of the exponential fit of the number of missing ordinal patterns to L [main text, Eq. (1)] for the AA time series (blue) and surrogate data (red). Panel (b): Box-plot diagrams comparing missing ordinal pattern decay time constant observed in the experimental AA time series data (blue) and the surrogate data (red). The black line in the center of the box represents the median. The notch represents 95% confidence interval of the median. The upper edge of the box represents the 75th percentile, and the lower edge of the box represents the 25th percentile. The upper whisker adds 1.5 times the inter-quartile range to the 75th percentile and the lower whisker subtracts 1.5 times the interquartile range from the 25th percentile. Circles denote the possible outliers. The median value of the rate of missing ordinal pattern decay in the experimental time series is lower than that of the surrogate data (Mann-Whitney test, p < 0.001), suggesting that the null hypothesis of AF dynamics being the result of a rescaled linear stochastic process is rejected.

When comparing the time constant of MOP decay in each individual time series against the time constant of MOP decay derived from the 40 IAAFT surrogate data corresponding to the original AA time series, the time constant of MOP decay was lower than the lowest 95% confidence interval band in 31/38 (81.6%) recordings (Fig. 6). On the 7/38 recordings that the time constant of MOP decay was not lower than the lowest 95% confidence interval band of the IAAFT MOP decay, the null hypothesis cannot be rejected. However, these recordings represent only a small proportion of our recordings.

### C. Causal entropy-complexity plane

Representation of the AA time series, IAAFT surrogate data, and fBm over the causal entropy-complexity plane is shown in Fig. 7. The colored curves represent the fitted least-squares cubic polynomial regression curves (blue is AA time series and green is fBm) and the interrupted colored curves the corresponding 95% confidence intervals of the AA time series and fBm. The IAAFT surrogate data have the same probability distribution with the original AA time series and thus, as expected, the same trajectory on the plane (marked as red dots on Fig. 7, fitted curve was omitted). The black interrupted lines demonstrate the maximal (upper line) and minimal (lower line) permutation statistical complexity possible for the corresponding permutation entropy levels and an embedding dimension of 5.[59] The coefficient of determination of the cubic models was 99.9% for both AA time series and fBm. The cubic function fitted to the AA time series showed a trajectory in higher complexity levels compared to that of fBm, with 99% confidence intervals that are mostly non-overlapping. The non-overlapping 99% confidence intervals suggest that the AA time series has a significantly higher complexity than fBm, suggesting that AA time series is a result of a deterministic or nonlinear stochastic process with >99% confidence. When a support vector machine classification approach was utilized to evaluate for discrimination between AA time series and fBm using the coordinates of each time series on the $\mathcal{H}[P] \times \mathcal{C}_{JS}[P]$ plane, 85% accurate classification was achieved (supplementary materials, Appendix C).

## IV. DISCUSSION

### A. Main findings

We demonstrate that human AF does not result from a rescaled linear stochastic process or a fractional noise. Our findings indicate that human AF results from a deterministic or a nonlinear stochastic process, rather than a rescaled linear stochastic process or a fractional noise. The MOP analysis using BP symbolization is robust to clinical time series that are inherently noisy[33–37] and can be applied even in the setting of irregular-sampling, missing data, and timing jitter.[45] The limitation of the MOP analysis to differentiate determinism from highly correlated noise is complemented by the causal entropy-complexity plane analysis[17,38] and the use of IAAFT surrogate data.[38,41] Analysis of the intracardiac electrograms with both methodologies yielded consistent results that human AF is deterministic or nonlinear stochastic with >99% confidence.

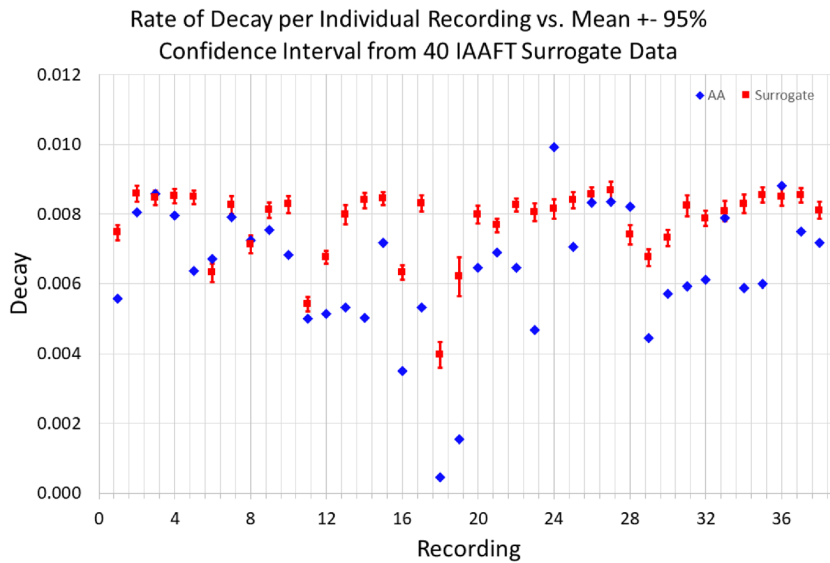

FIG. 6. Missing ordinal pattern decay time constant for each individual recording compared to the mean ±95% confidence intervals of missing ordinal pattern decay time constant of their corresponding 40 IAAFT surrogate data. The X-axis represents consecutive recordings (1–38) and the y-axis the missing ordinal pattern decay time constant that is calculated as described in the main text and depicted in Fig. 5(a). Blue diamonds represent the missing ordinal pattern decay time constant of each AA time series. Red squares represent the mean missing ordinal pattern decay time constant of the 40 IAAFT surrogate datasets corresponding to the same AA time series. The whiskers represent 95% confidence intervals. The time constant of missing ordinal pattern decay was lower than the lowest 95% confidence interval band in 31/38 (81.6%) recordings.

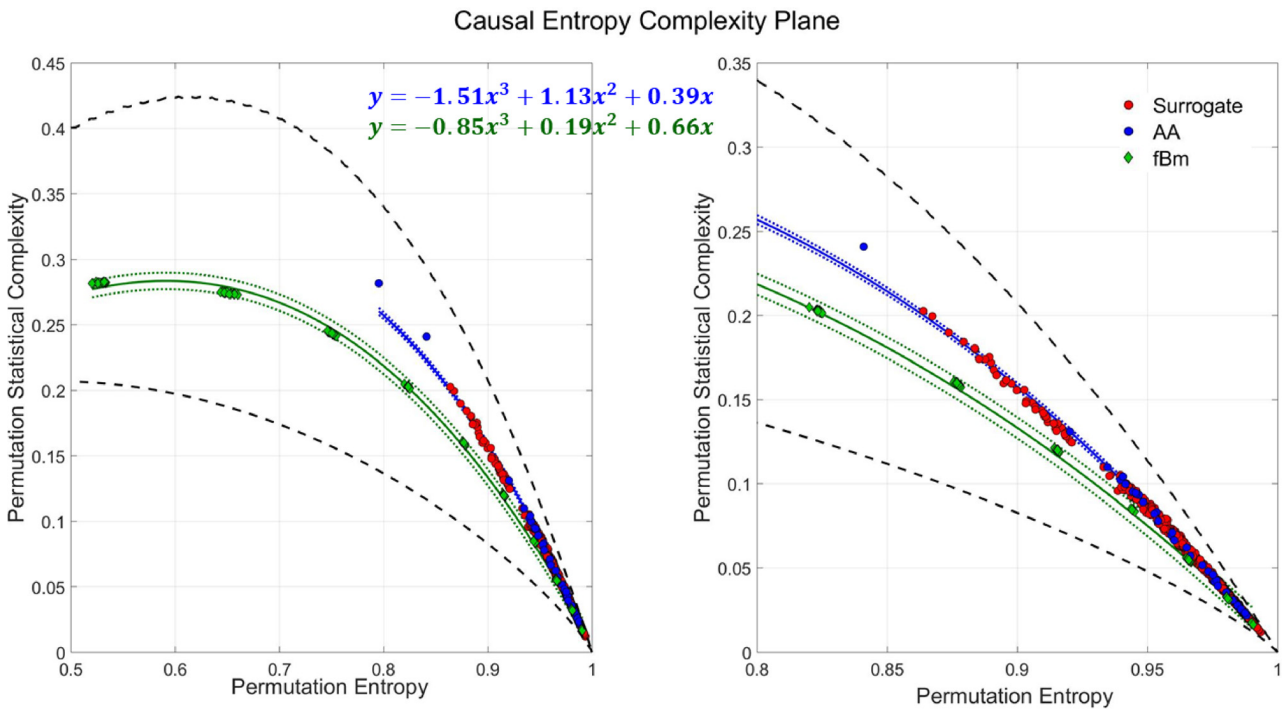

FIG. 7. Causal entropy-complexity plane. The X-axis represents permutation entropy and the Y-axis represents permutation statistical complexity. Each time series is represented with a point on this plane. Experimental AA time series are shown in blue, surrogate data are shown in red, and fractional Brownian motion with the increasing Hurst parameter (0 < H < 1) is shown in green. The colored curves represent best-fit cubic curves to the points, and the interrupted lines represent the 99% confidence intervals (blue is AA time series and green is fractional Brownian motion). As the Hurst exponent decreases from 1 to 0.1, the points of fractional Brownian motion move towards the right lower corner of the plane (maximal entropy with minimal complexity). The interrupted black lines represent the maximal and minimal values of permutation statistical complexity which are possible for the corresponding levels of permutation entropy. The equations represent the polynomials describing the curves fitted to the AA time series (blue) and fractional Brownian motion (green). As expected, since the surrogate data have the same probability distribution as the AA time series, their trajectory completely overlaps with that of the AA time series. The panel on the right focuses on the lower left corner of the entropy complexity plane, for better visualization of the points and confidence intervals that correspond to the highest entropy and lowest statistical complexity values. The AA time series have a trajectory that is significantly above the trajectory of the fractional Brownian motion. The null hypothesis that AF is the result of a fractional noise is thus rejected.

### B. Determinism of atrial fibrillation

To our knowledge, this is the first study to demonstrate that human AF is not the result of a rescaled linear stochastic process or a fractional noise using the BP methodology. The MOP analysis has been used to characterize the dynamics of heart rate variability[60] which reflects the autonomic function rather than cardiac dynamics, but it has never been applied to human AF. Our findings help to improve our understanding of human AF at multiple levels. For example, our results validate the effort to develop a deterministic or a nonlinear stochastic model to simulate AF.[6–10] In addition, our results justify the use of nonlinear dynamical tools to describe AF properties.

In light of the literature evaluating determinism in human AF, our work is highlighted by the application of novel methodology to address the limitations of earlier studies. To date, there are only three studies available that directly evaluated the determinism of human AF, and their conclusions were conflicting. Two of those studies (n = 5 and 7) found determinism in AF[13,15] and the other study failed to demonstrate determinism (n = 9).[14] In those earlier studies, the methods to detect determinism in human AF included Poincaré plot analysis,[13] Grassberger-Procaccia correlation dimension, correlation entropy, coarse-grained correlation dimension and coarse-grained correlation entropy,[14] Lyapunov exponent, Kolmogorov entropy, and Lempel-Ziv complexity.[15] All of those methods are limited by the sensitivity to experimental noise, low robustness with shorter durations of time series, and the sensitivity to initial parameter selection.[25–31] In addition, the Grassberger-Procaccia method and Lyapunov exponents[14,15] could falsely classify highly correlated stochastic time series as deterministic.[25,32] The critical strength of our work is that we used a combination of the BP symbolization with the Amigó methodology and the causal entropy-complexity plane, both of which are robust against all of those limitations.[33–37,45] Importantly, the two separate methods showed consistent results. Another strength of our work is that we used the IAAFT surrogate data. Only one of the earlier studies described above used a surrogate data framework.[14] The use of surrogate data is critical for accurate evaluation of time series that have the potential to be contaminated by noise.[61] The IAAFT surrogate data have been used with the BP MOP paradigm successfully, in both theoretical and experimental settings.[41] We used the causal entropy-complexity plane of the BP symbolization of the AA time series and the fBm to differentiate deterministic time series from highly correlated noise.[17,38] In addition, we used rigorous statistical analysis to quantify our findings using a relatively large sample size of AF episodes.

Finally, several studies provide indirect evidence of determinism in AF. For example, simulation studies suggest that AF may arise through a quasiperiodic transition to chaos,[13] and conversion from AF to atrial flutter is a phase transition.[16] In human AF, spatiotemporal organization of atrial electrical activity has been demonstrated.[62,63] In addition, human AF had higher values of several nonlinear parameters such as Lyapunov exponent, Kolmogorov entropy, and Lempel-Ziv complexity compared to human typical atrial flutter.[15] However, an additional strength of our work is that it evaluates real-life, clinically acquired electrophysiology recording that directly capture AF dynamics.

### C. Limitations

This study has several limitations. First, our findings may be applicable only to the location of the catheter-based intracardiac recordings of AF. For example, our data derived exclusively from the coronary sinus, which is anatomically adjacent to the inferior aspect of the left atrium. It is possible that intracardiac measurements from other parts of the left atrium or the right atrium may have led to a different result. However, we chose to use the electrograms from the coronary sinus because its anatomical structure allows persistent stabilization of the measurement catheter for the entire duration of measurements to minimize motion-induced noise in the beating human heart. Second, the MOP and causal entropy-complexity plane analysis presented herein are not designed to discriminate between high dimensional and low dimensional deterministic dynamics. The MOP approach can detect determinism even in high-dimensional dynamical systems.[36] Traditional tools of low-dimensional non-linear dynamics are not directly applicable to the analysis of high-dimensional dynamical systems. Demonstration of determinism however is still relevant, as utilization of low-dimensional descriptions of high-dimensional systems can be feasible and is currently is an active field of investigation. Third, neither the MOP paradigm nor the causal entropy-complexity plane analysis has been validated in discriminating determinism against nonlinear stochastic processes, and thus, the results presented herein cannot ascertain that AF is not the result of such a process. Fourth, rejection of our null hypothesis does not necessarily imply non-linear dynamics. For instance, non-instantaneous measurement functions can lead to rejection of the null hypothesis, although the underlying dynamics may be linear. However, there is no evidence or theories to date to suggest that bipolar voltage or AA intervals would represent such a measurement function. Further, the aim of this work was to test the hypothesis that human AF is not the result of a rescaled linear stochastic process or a fractional noise, rather than to provide an exhaustive description of its dynamical properties, which would require an extended time series, an acquisition of which is not feasible in the current clinical practice. Fifth, separation of the causal entropy-complexity plane in a "deterministic" and a "stochastic" area using the trajectory defined by fBm is arbitrary and there is no analytical proof of the generalizability of this statement. It is however an acceptable boundary for detection of determinism in limited studies that utilize the causal entropy-complexity plane for detection of determinism in experimental time series.[56,57] Finally, the sample size of this study is relatively small considering all cases of AF in the population. Thus, the generalizability and external validity of our results to all cases of AF might be limited. However, the sample size of this study is the largest amongst other studies assessing for determinism in AF to date.

### D. Conclusions

Analysis of human AF using missing ordinal patterns and the causal entropy-complexity plane of the Bandt-Pompe symbolization in a surrogate data framework suggests that human AF is not driven by a rescaled linear stochastic process or a fractional noise. Our results support the development and application of mathematical analysis and deterministic or nonlinear stochastic modeling tools to enable predictive control of human AF.

### SUPPLEMENTARY MATERIAL

See supplementary material for a detailed description of the Band-Pompe methodology *(Appendix A)*, calculation of permutation entropy and permutation statistical complexity *(Appendix B)*, and the support vector machine approach that we used for discrimination between fBm and AA on the causal entropy-complexity plane *(Appendix C)*.

### ACKNOWLEDGMENTS

The research reported in this publication was supported by the National Heart Lung and Blood Institute of the National Institutes of Health under Award No. T32HL007227 (to K.N.A.). The percentage of the total program project costs financed with Federal money is 100%. The content is solely the responsibility of the authors and does not necessarily represent the official views of the National Institutes of Health. This research was also supported by the Fondation Leducq Transatlantic Network of Excellence 16CVD02 (to H.A.). The authors have no conflict of interest to disclose.